# Temperature dependent reversal of voltage modulated light emission and negative capacitance in AlGaInP based multi quantum well light emitting devices

Kanika Bansal and Shouvik Datta

Division of Physics, Indian Institute of Science Education and Research,

Pune - 411008, Maharashtra, India

## Abstract

We report a reversal in negative capacitance and voltage modulated light emission from AlGaInP based multi-quantum well electroluminescent diodes under temperature variation. Unlike monotonically increasing CW light emission with decreasing temperature, modulated electroluminescence and negative capacitance first increase to a maximum and then decrease while cooling down from room temperature. Interdependence of such electronic and optical properties is understood as a competition between defect participation in radiative recombination and field assisted carrier escape from the quantum well region during temperature variation. The temperature of maximum light emission must coincide with the operating temperature of a device for better efficiency.

Improvement in the performance efficiency of electroluminescent diodes (ELDs) is the primary goal of physicists as well as engineers working in the field of optoelectronics. Continuous research is going on to achieve better efficiencies for a vast variety of technological applications[1]. For efficiency improvement as well as from basic understanding point of view, it is necessary to study the interdependence of electronic and optical properties of ELDs under their operating condition of light emission. With this motivation we studied steady state behavior of commercially available red-ELDs (AlGaInP based multi-quantum well laser diodes) during light emission below the lasing threshold. Conventional characterization techniques based on depletion approximation and electrostatic description of the junction may not work within their usual interpretation in this high forward bias regime beyond[2] the range of diffusion capacitance.

We probed the steady state dielectric response of the ELDs by studying both voltage modulated electroluminescence[2] (VMEL) and junction impedance. In our earlier work[2] with these diodes (for frequencies ≤ 100 kHz) at room temperature we showed that under sufficiently high forward bias and low frequency, reactive component of the impedance acquires 'inductive like' behavior commonly referred as negative capacitance (NC), which is exciting, technologically important[3,4] and yet less understood phenomenon. Interestingly our VMEL results showed[2] that the modulated light emission is closely related to NC and both depend systematically on applied modulation frequencies. Frequency dependent NC and VMEL were understood as the outcome of dynamic competition between radiative recombination process in the quantum well (QW) and the steady state response from shallow defects. This produces a transient change in the charge carrier reservoir available for recombination and triggers a compensatory inductive like response[2]. In the work we assumed that at constant room temperature steady state population of charge carriers inside the QW is unaffected by any escape of carriers over the QW barrier. Results presented in this letter also support this assumption. The observed low frequency response can be a cause of efficiency reduction during high frequency direct modulation[2] (~GHz) used in applications like optical communication. This is because defect mediated steady state contribution of carriers undergoing radiative recombination will be missing in action at such high frequency. Hence we had emphasized the need to carry out such low frequency steady state characterizations to optimize the device design for better light emission efficiency.

Under low frequency modulation, NC is observed by other groups in ELDs[5-7] and also in different semiconductor devices[8]. Time domain study of frequency dependent NC has also been done[9-11]. However for the first time, to our knowledge, we are reporting the temperature variation of both NC and VMEL and their interdependence. Unlike usual[12] continuous wave (CW) light emission which expectedly decreases continuously with increasing temperature, both NC and VMEL go through a maximum during temperature scan. Position of this maximum systematically shifts towards higher temperature with increasing forward bias. This occurrence of peak or maxima in modulated light emission immediately indicates that the device can have maximum efficiency at different temperatures for different biases. Thus, we suggest that for efficient working of ELDs under direct modulation, the temperature of maximum light emission must be tailored to their operational temperature. This can be achieved either by careful choice of materials and/or by clever device engineering.

Fig. 1a shows the variation of VMEL or modulated light emission for commercially available laser diode (Sanyo, DL3148-025) with temperature at modulation frequency of 1 kHz under different current levels below lasing threshold. Experiments were done at constant forward bias current ($I_{dc}$), which allows us to ignore any effect of change in the total charge carrier density on the mechanisms we are probing at various temperatures (accuracy of ± 0.5K). Details of the experimental setup and devices can be found in ref. 2. For a given $I_{dc}$, when the sample is cooled down from room temperature, VMEL signal increases and interestingly, within a certain range of temperatures (~140K – 210K), reverses the nature and starts to decrease giving a maximum in light emission. This temperature of maximum emission ($T_{Max}$), shifts systematically to higher temperature side as the bias increases. However CW light emission (fig. 1b) expectedly increases with cooling. In general, efficiency of radiative recombination improves with decreasing temperature in QW because of the decreased radiative recombination lifetime[13]. Absence of any maximum in the case of CW light emission indicates that the observed features in VMEL are only due to steady state response of charge carriers to applied voltage modulation. This is evident in fig. 1c which shows the frequency variation of the temperature dependent VMEL measured at $I_{dc}$ = 250 μA. For any given temperature, lower frequency produces higher VMEL signal which is in agreement with our earlier reported results[2]. Noticeably, $T_{Max}$ is relatively independent of the frequency (for frequencies ≤ 100 kHz). Figures 2a and 2b show temperature variation of capacitance and conductance respectively for different injection levels at a modulation frequency of 1 kHz. These measurements are done under the circuit model containing capacitance and conductance in parallel as described in ref. 2. In capacitance also we observed similar temperature dependent reversal giving a temperature ($T_{Max}$) where NC is maximum. Here also $T_{Max}$ shifts to higher temperature with increasing forward bias. Similar changes are seen in conductance too. Fig. 2c shows that like VMEL, $T_{Max}$ is independent of modulation frequency in case of NC too. NC is more prominent for lower frequencies for all the temperatures. Figures 1 and 2 certainly indicate the correlation between measured optical (VMEL) and electronic (NC) properties. We will explain this correlated behavior in terms of active participation of defects in radiative recombination dynamics and continuous escape of charge carriers from the quantum well region.

Defect states can only contribute to the VMEL and NC signal if their thermal rate of carrier exchange with the band edges ($1/\tau$) is comparable to applied frequency ($f$). Here ($1/\tau$) is related to the thermal activation energy ($E_{Th}$) of the defect level by $1/\tau = \nu \exp(-E_{Th}/k_B T)$ where $k_B$ is the Boltzmann constant, $\nu$ is the thermal prefactor, hence we can write:

$$E_{Th} = k_B T \ln(\nu / f). \qquad (1)$$

Steady state charge carrier (e.g. electrons) contribution from defects is given as $n^{Defects}(E) = \int_{E_C - E_{Th}}^{E_{Fn}} g(E)dE$, g(E) is the sub-bandgap defect density. Limits of integration represent[2] the total defect states which can respond to the applied signal as given by equation 1 and also by the position of quasi Fermi level ($E_{Fn}$) and band edges (conduction band $E_C$ in case of electrons). Interestingly, at a given bias and frequency, defect contribution to charge carrier density available for radiative recombination increases with the increasing temperature. This is why we see counterintuitive increase in the magnitude of VMEL and NC signal as shown by figures 1a and 2a, when temperature is increased from a quite low value (~ 60 K). However this increase is rather slow for relatively low temperatures (≤ 100 K) due to the inability of charge carriers to respond to the applied voltage modulation. In such case one would expect that for even lower temperatures both VMEL and NC will saturate. However, with further heating VMEL and NC reverse the behavior and start to decrease above certain temperature (> $T_{Max}$). Likely cause is the increasing loss of charge carriers available for radiative recombination due to their field assisted thermal escape from the active QW region. This is when the QW escape process dominates over shallow defect mediated contribution to steady state radiative recombination. In addition to this escape, typical activation of deep defects with very large response time may also cause the net loss of charge carriers in the QW within the measurement time window. Schematic of these processes is shown in fig. 3a. It is interesting to note that thermally activated QW escape rate and consequently the $T_{Max}$ seems to be independent of modulation frequency within our range of measurements. In fact, the independence of $T_{Max}$ within the range of applied modulation frequency (fig. 1c and 2c) favors the presence of very fast QW escape processes around room temperature to explain the reversal observed in VMEL and NC.

To probe the escape process we measured the variation of CW light emission with temperature for different $I_{dc}$ values (fig. 3b) and also the variation in $I_{dc}$ with temperature while maintaining certain light emission intensity (fig 3c). Continuous charge carrier escape from QW is a complimentary process to light emission in a sense that it depletes the number of charge carriers available for radiative recombination. In our case, a major contribution to this $I_{dc}$ comes from charge injection to active region, QW escape, radiative recombination and trapping by defects etc. Current density of escape $J_E$ for electrons in a QW is[14]

$$J_E = \frac{n_{QW} e L}{\tau_E} \qquad (2)$$

where $L$ is the width of the QW, $n_{QW}$ is the carrier density inside the well, $e$ is the electronic charge and $1/\tau_E$ is the rate of thermal escape. As the temperature increases, $1/\tau_E$ for charge carriers in the QW increases, and escape process starts to dominate. This is also evident in fig. 3b where falling rate of radiative recombination is the outcome of increasing temperature at a fixed $I_{dc}$ due to higher consumption rate of charge carriers by the thermally activated QW escape process. At a particular temperature more light emission is observed for higher $I_{dc}$ as expected.

Similarly in fig. 3c, as the temperature increases, higher rate of charge injection in terms of $I_{dc}$ is required to maintain constant light emission intensity. This is again due to the increase in thermally activated QW escape process with increasing temperature. At a particular temperature, required $I_{dc}$ to maintain light emission increases with increasing light emission intensity as is also discussed in the context of fig. 3b. Straight line behavior in Arrhenius like plots of figures 3b and 3c show thermal activation rate of light emission and injected charge respectively. Calculated values of respective activation energies $E_I$ and $E_{II}$ are shown in the insets. The rate for any thermally activated QW escape process can be given as[13]

$$\frac{1}{\tau_E} = \left(\frac{k_B T}{2\pi Q}\right) \exp\left(-\frac{E_B}{k_B T}\right) \tag{3}$$

where $Q$ is the factor related to $L$ and effective mass of the charge carrier inside the well and $E_B$ is the effective thermal barrier height. In the present case of active light emitting device under voltage modulation, thermal rate of one process is intricately connected with the other competing processes like charge injection, recombination, escape and defect trapping etc. Many of these processes are also intrinsically thermally activated with different barrier heights. Hence it is difficult to directly associate the calculated activation energies solely to any particular barrier height of field assisted thermal escape from QW. However we understand $E_I$ as the activation energy of light emission which goes down with increasing bias ($\sim I_{dc}$) suggesting that the QW escape process which is complimentary to light emission decreases with increasing $I_{dc}$. This happens due to the field assisted behavior of QW escape process which is also evident in figures 1a and 2a. In absence of any forward bias, built-in field of the p-n junction efficiently separates thermally escaped charge carriers[15], however with increasing forward bias ($I_{dc}$) this net field contribution to field assisted thermally activated escape of carriers from QW decreases. As a result $T_{Max}$ shifts towards higher temperature for higher $I_{dc}$ as can be seen in figures 1a and 2a. $E_{II}$ is understood as the activation energy of charge injection which remains relatively unchanged with light emission intensity within our range of measurements and mainly depends on p-n junction parameters of ELD. To sum up, as the temperature increases, field assisted thermal escape process gets strengthened and dominates over defect mediated radiative recombination process, consequently we see a decrease in the magnitude of VMEL and NC signals (figures 1a and 2a) above a certain temperature $T_{Max}$.

In conclusion, we report counter intuitive decrease in the magnitude of both VMEL and NC from red electroluminescent diodes below a temperature $T_{Max}$ which deviates from the usual monotonically increasing nature of light emission with decreasing temperature. Here $T_{Max}$ is the temperature where modulated light emission is maximum. This is caused by the interplay of active participation of shallow defect levels in radiative recombination and the continuous charge carrier escape processes from the quantum well active region. This $T_{Max}$ systematically shifts to higher temperature side with increasing forward bias and is relatively independent of modulation frequency within the range of our measurements. Similar qualitative features are also observed in negative capacitance. This correlated behavior of optical and electronic processes is important to understand the device behavior during modulated light emission. From application point of view, the temperature of maximum light emission must be brought to operating temperature of the light emitting device for its efficient performance under direct modulation. Although here we presented results on AlGaInP based multi quantum well laser diodes below lasing threshold, we expect these results and analyses to be valid for a wider range of light emitting devices. To further understand the correlation between observed results, a rigorous theoretical frame work beyond depletion approximation and electrostatic description of the junction is required along with the experiments for varying quantum well parameter, defect dynamics and modulation parameters.

Authors wish to thank IISER-Pune for startup funding of the laboratory infrastructure and Dept. of Science and Technology, India for DST Nano Unit grant SR/NM/NS-42/2009. KB is thankful to CSIR, India for research fellowship.

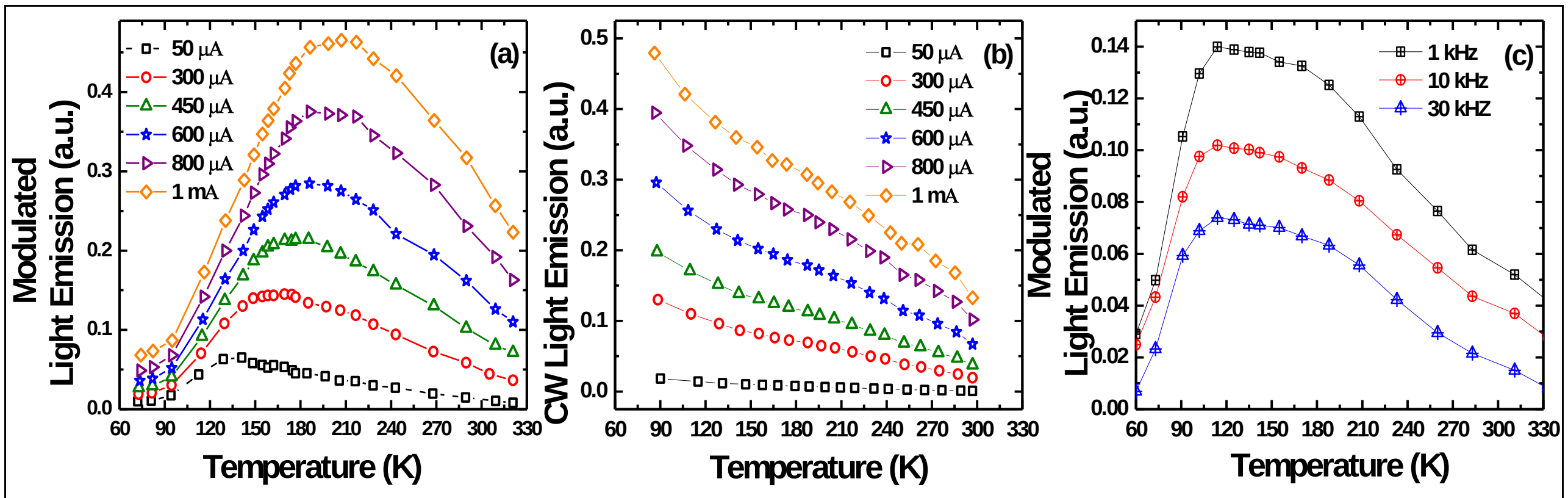

Figure 1: Temperature scan of light emission from laser diode at different $I_{dc}$. (a) Voltage modulated light emission at 1 kHz showing maxima at certain temperatures ($T_{Max}$). Increasing bias shifts $T_{Max}$ to higher temperature. For cost effective performance, this maximum should be around operational temperature of the ELD. (b) Usual CW light emission showing a continuous increase with decreasing sample temperature. (c) Frequency dependence of temperature scan of modulated light emission at 250 μA. Note that $T_{Max}$ is relatively independent of applied modulation frequencies.

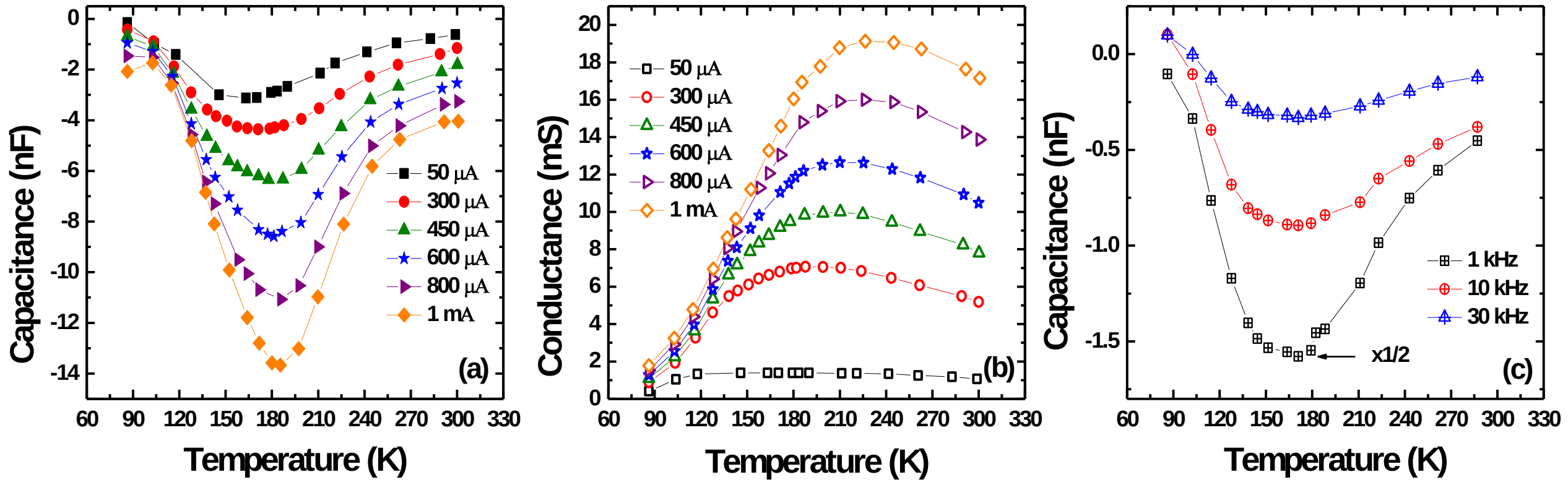

Figure 2: Temperature scan of (a) negative capacitance and (b) conductance for laser diode at 1kHz for different $I_{dc}$. A maximum is observed at certain temperature ($T_{Max}$) which shifts to higher side for higher bias. (c) Frequency dependence of the temperature scan of negative capacitance at 250 μA. Again $T_{Max}$ is relatively independent of applied modulation frequency.

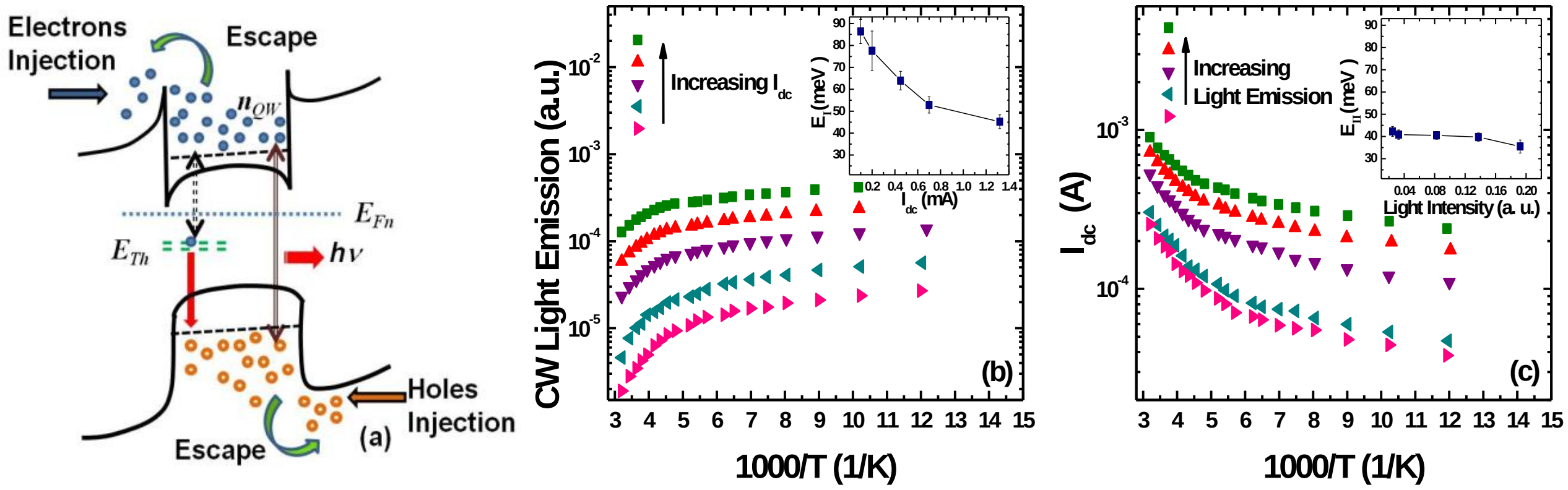

Figure 3: (a) Schematic band diagram of the electronic processes occurring in QW. Electrons (holes) are injected from the n (p) neighborhood of the junction. These electrons (holes) can undergo radiative recombination or defect mediated recombination through some defect with activation energy $E_{Th}$. With increasing temperature, field assisted thermal escape of charge carriers starts to dominate, thereby reducing the number of charge carriers inside the well ($n_{QW}$). $E_{Fn}$ shows the quasi Fermi level for electron within QW for small forward bias. Quasi Fermi level for holes is not shown here for simplicity. (b) Showing increase in the rate of light emission with decreasing temperature and increasing $I_{dc}$. Absolute value of the activation energy for light emission $E_I$ as calculated from the straight line portion of the Arrhenius plot is given in the inset. (c) Variation in the rate of charge injection in the form of $I_{dc}$ with temperature to maintain certain CW light emission intensity. Increasing light emission requires higher $I_{dc}$. Calculated activation energy for charge injection $E_{II}$ is shown in the inset.